\documentclass[journal]{IEEEtran}
\usepackage{amssymb}
\usepackage{makecell}
\usepackage{multirow}
\usepackage{graphicx}
\usepackage{cite}
\usepackage{bm}
\usepackage{pifont}
\usepackage{bbding}
\usepackage{amsmath}
\usepackage{amssymb}
\usepackage{mathrsfs,balance}
\usepackage{algorithm}
\usepackage{algorithmic}
\usepackage{color}
\usepackage{amsthm}
\usepackage{booktabs}
\usepackage{stfloats}

\begin{document}
\title{Uplink Networked Sensing via Multiuser Correlation Exploitation}

\author{
        Jingying~Bao,
        J. Andrew~Zhang,~\IEEEmembership{Senior Member,~IEEE},
        Kai~Wu,~\IEEEmembership{Member,~IEEE},
        Christos~Masouros,~\IEEEmembership{Fellow,~IEEE},
        and~Y. Jay~Guo,~\IEEEmembership{Fellow,~IEEE}

\thanks{J. Y. Bao, J. A. Zhang, K. Wu and Y. J. Guo are with the Global Big Data Technologies Centre, University of Technology Sydney, Sydney, NSW 2007, Australia (e-mail: Jingying.Bao@student.uts.edu.au; andrew.zhang@uts.edu.au; kai.wu@uts.edu.au; jay.guo@uts.edu.au).}
\thanks{C. Masouros is with the Department of Electronic and Electrical
Engineering, University College London, London WC1E 7JE, U.K. (e-mail:
c.masouros@ucl.ac.uk)}
}
\maketitle
\begin{abstract}
In this correspondence, we investigate networked sensing in perceptive mobile networks under a bistatic multi-transmitter single-receiver uplink topology, where multiple user equipments (UEs) transmit signals over orthogonal frequency-division multiple access (OFDMA) resources and a single base station performs joint sensing. Uplink clock asynchronism introduces offsets that destroy inter-packet coherence and hinder high-resolution sensing, while multi-user observations exhibit exploitable cross-user correlation. We therefore formulate an asynchronous multi-user uplink OFDMA sensing model and exploit common delay-cluster sparsity across UEs. A line-of-sight (LoS)-referenced calibration first suppresses the offsets, after which a shared-private delay-domain sparse Bayesian learning (SBL) model is used for delay support recovery and user grouping. Doppler and angle of arrival are then estimated from temporal and spatial phase differences. Simulation results show that the proposed scheme outperforms per-user processing, particularly under limited subcarrier budgets and in low signal-to-noise ratio (SNR) regimes.
\end{abstract}
\section{Introduction}
Networked sensing is a key capability of perceptive mobile networks (PMNs) for integrated sensing and communication (ISAC), where distributed connected nodes cooperate to acquire environmental information at scale \cite{9296833}. By leveraging multi-perspective observations across  geographically separated transceivers (e.g., multi-site/cooperative architectures), networked sensing can improve sensing coverage, robustness, and estimation accuracy over isolated sensing \cite{10557715,10735119}. However, realizing such gains typically requires effective coordination and multi-node fusion under practical network constraints, posing implementation challenges in PMN/ISAC \cite{10077114}.

Existing studies on networked sensing in PMNs/ISAC mainly focus on multi-site cooperation or multi-receiver architectures, where sensing gains are achieved by combining observations from distributed sites/base stations (BSs) \cite{10494224,10304081}. For example, Behdad \emph{et al.} in \cite{10494224} investigated multi-static target detection in cell-free massive multiple-input multiple-output via cooperation among distributed access points, while Li \emph{et al.} in \cite{10304081} studied cellular multistatic ISAC for seamless sensing coverage via cooperation among spatially separated nodes. However, realizing such cooperation in practice typically requires tight inter-site coordination, including accurate synchronization and coordinated interference management, thereby incurring considerable system complexity and signaling overhead.

To this end, we consider a novel bistatic multi-transmitter single-receiver (multi-Tx/single-Rx) topology for uplink networked sensing, where multiple user equipments (UEs) transmit to one BS, consistent with most practical communication networks, including cellular and Wi-Fi uplinks. Compared with multi-site cooperative sensing, this paradigm-shifting topology avoids inter-site coordination while still benefiting from multi-user diversity, where UEs in similar propagation environments often exhibit correlated multipath signals reflected/defracted from common targets \cite{8552436}. One of the major challenges here is how to exploit such correlation as conventional sensing algorithms such as multiple signal classification (MUSIC) lacks such a capability. Another challenge is bistatic asynchronism, which introduces UE-dependent and time-varying timing offset (TO), carrier-frequency offset (CFO), and phase offset (PO), thereby degrading inter-packet coherence and hindering high-resolution sensing \cite{9349171}. 

To address these issues, we develop an asynchronous uplink networked sensing framework by introducing  compressive sensing techniques to explore multiuser correlation, in combination with offsets cancellation techniques due to bistatic asynchronism. The framework is demonstrated via exploiting common delay cluster only, but it can be naturally extended to other sensing parameters, such as Doppler and angles, and correlation patterns. Specifically, we first perform LoS-referenced calibration to suppress the dominant TO and align packets to a common delay reference. Then, we propose a shared-private delay-domain sparse Bayesian learning (SBL) formulation that jointly recovers delay supports while automatically exploiting correlation. Finally, based on the recovered delay taps, we compensate CFO/PO and then estimate Doppler and angle of arrival (AoA). Simulation results show that exploiting common delay-cluster sparsity outperforms the individual-SBL baseline, especially at low signal-to-noise ratio (SNR) and with small per-UE subcarrier budgets.

\begin{figure}[t]
\centering
\includegraphics[scale=0.4]{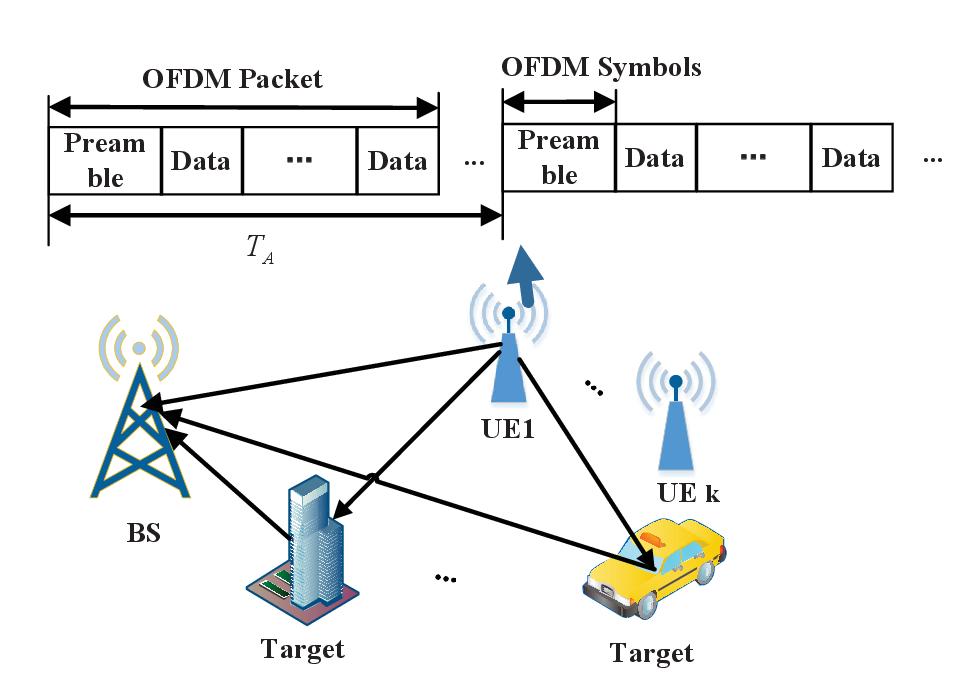}
\caption{Illustration of the system model for uplink sensing.}
\label{fig:model}
\end{figure}

\section{System Model}
\subsection{Signal Model }
We consider an uplink ISAC scenario in a PMN as shown in Fig. 1, where $K$ single-antenna UEs transmit to a BS equipped with an $M$-element uniform linear array (ULA) in an uplink orthogonal frequency-division multiple access (OFDMA) system.  UE $k$ occupies a dedicated subcarrier set $\mathcal{N}_k\subset\{1,\ldots,N\}$ with $N_k\triangleq|\mathcal{N}_k|$, where $N$ is the total number of subcarriers, each UE is assigned a contiguous block of $N_k$ subcarriers and different UEs are allocated non-overlapping blocks (i.e., $\mathcal{N}_k\cap\mathcal{N}_{k'}=\emptyset$ for $k\neq k'$). The BS collects $T$ OFDM packets indexed by $t\in\{1,\ldots,T\}$. In particular, we assume that the BS knows the distance from each fixed UE, hence the LoS propagation delay $\tau_{k,0}^{\rm geom}$ is known, and the LoS path power is significantly stronger than that of the NLoS paths. This provides a delay reference for uplink sensing and synchronization, while the remaining non-line-of-sight (NLoS) components are to be estimated. These assumptions are reasonable and commonly adopted in PMN/ISAC uplink sensing \cite{9349171,10966468}.

As depicted in Fig.~1, each uplink communication packet consists of one reference-bearing preamble and a sequence of data symbols with OFDM modulation, and two consecutive packets are spaced by $T_A$. 
For simplicity, each UE is assumed to use only one preamble OFDM symbol for sensing in each packet. In addition, due to the asynchronous transceivers, the received signal may be corrupted by the packet-dependent TO/CFO and PO, denoted by $\delta\tau_k[t]$, $\delta \nu_k[t]$, and $\beta_k[t]$, respectively. As a result, for UE $k$ at packet $t$, after pilot matched filtering, the channel state information (CSI) estimation can be written in the delay-domain form as
\begin{equation}
\mathbf{Y}_k[t]
=
\sum_{\ell=1}^{L_k}
\boldsymbol{\psi}_k\!\left(\tilde{\tau}_{k,\ell}[t]\right)
\mathbf{s}_{k,\ell}^{\top}[t]
+
\mathbf{E}_k[t],
\label{eq:Y_sum}
\end{equation}
where $L_k$ denotes the number of paths for UE $k$,
\[
\boldsymbol{\psi}_k(\tilde{\tau})
\triangleq
\big[
e^{-j2\pi f_{n_1}\tilde{\tau}},
\ldots,
e^{-j2\pi f_{n_{N_k}}\tilde{\tau}}
\big]^{\top}\in\mathbb{C}^{N_k}
\]
is the delay steering vector over $\mathcal{N}_k$ and $\tilde{\tau}_{k,\ell}[t]\triangleq \tau_{k,\ell}+\delta\tau_k[t]$;
$\mathbf{s}_{k,\ell}[t]\triangleq \alpha_{k,\ell}e^{j\beta_k[t]}e^{j2\pi tT_A(\nu_{k,\ell}+\delta\nu_k[t])}\mathbf{a}(\theta_{k,\ell})$ is the effective spatial coefficient; $\{\alpha_{k,\ell},\, \tau_{k,\ell},\, \nu_{k,\ell},\, \theta_{k,\ell}\}$ denote the sensing parameters; $\mathbf{a}(\theta_{k,\ell}) =\big[1,\; e^{j\pi \sin(\theta_{k,\ell})},\; \ldots,\; e^{j\pi (M-1)\sin(\theta_{k,\ell})}\big]^{\top}\in\mathbb{C}^{M}$ is the receive array response vector of the ULA, and
$\mathbf{E}_k[t]\in\mathbb{C}^{N_k\times M}$ denotes the additive white Gaussian noise matrix.

\subsection{Proposed Sparsity Model Capturing Delay Correlation}
To facilitate delay estimation, we discretize the effective delay domain into a common uniform grid
$\{\bar{\tau}_1,\ldots,\bar{\tau}_G\}\subset[0,\tau_{\max}]$,
where $G$ is the number of grid points, and construct 
\begin{equation}
\mathbf{\Psi}_k \triangleq
\big[\boldsymbol{\psi}_k(\bar{\tau}_1),\ldots,\boldsymbol{\psi}_k(\bar{\tau}_G)\big]
\in\mathbb{C}^{N_k\times G}.
\label{eq:Psi_k}
\end{equation}

Accordingly, by approximating $\tilde{\tau}_{k,\ell}[t]$ to its nearest grid point in
$\{\bar{\tau}_g\}_{g=1}^G$, \eqref{eq:Y_sum} can be simplified as the following multiple-measurement-vector (MMV) model
\begin{equation}
\mathbf{Y}_k[t]=\mathbf{\Psi}_k\,\mathbf{W}_k[t]+\mathbf{E}_k[t],
\label{eq:mmv_basic}
\end{equation}
where $\mathbf{W}_k[t]=[\mathbf{w}_{k,1}[t],\,\mathbf{w}_{k,2}[t],\,\ldots,\,\mathbf{w}_{k,G}[t]]^{\top}$ is a row-sparse coefficient matrix on the delay grid, $\mathbf{w}_{k,g}^{\top}[t] \triangleq \sum_{\ell:\, g_{k,\ell}=g} \mathbf{s}_{k,\ell}^{\top}[t]$ and $\mathbf{w}_{k,g}^{\top}[t]=\mathbf{0}$ if no path is assigned to $\bar{\tau}_g$. Hence, estimating delay reduces to recovering the row support of $\mathbf{W}_k[t]$.\footnote{Since $\mathbf{\Psi}$ is parameterized by delay alone, Doppler/CFO/PO and AoA are absorbed into $\mathbf{w}_{k,g}^{\top}[t]$ and only affect the values of the nonzero coefficients, without changing the active delay indices.}

However,  \eqref{eq:mmv_basic} ignores structured sparsity across UEs.
In clustered multi-user ISAC/OFDMA channels, after excluding the TO-induced $\delta\tau_k[t]$,  UEs may exhibit cross-user commonality in the sensing parameters under similar propagation conditions. Motivated by this observation, we form clusters based on delay similarity, which leads to partially common delay supports among UEs in the same cluster. To this end, we introduce the following shared--private decomposition \cite{8552436}:
\begin{equation}
\mathbf{W}_k^{\rm nTO}[t]=\mathbf{W}^{\rm sh}_k[t]+\mathbf{W}^{\rm pr}_k[t],
\label{eq:W_sum_new}
\end{equation}
where $\mathbf{W}_k^{\rm nTO}[t]$ denotes the sparse delay-domain coefficient matrix after TO removal, $\mathbf{W}_k^{\mathrm{sh}}[t]$ collects the coefficients on the delay taps commonly shared by UEs in the same cluster, while $\mathbf{W}_k^{\mathrm{pr}}[t]$ captures user-specific components. As a simple example, if two UEs in one cluster have active delay supports $\{\tau_2,\tau_5,\tau_9,\tau_{14}\}$ and $\{\tau_2,\tau_5,\tau_9,\tau_{20}\}$, respectively, then the common taps $\{\tau_2,\tau_5,\tau_9\}$ are represented by $\{\mathbf{W}_1^{\mathrm{sh}}[t],\mathbf{W}_2^{\mathrm{sh}}[t]\}$, whereas $\tau_{14}$ and $\tau_{20}$ are modeled by $\mathbf{W}_1^{\mathrm{pr}}[t]$ and $\mathbf{W}_2^{\mathrm{pr}}[t]$, respectively. This decomposition allows common supports to be reinforced jointly across UEs while preserving individual flexibility, thereby improving support recovery.

Therefore, after removing $\delta\tau_k[t]$, the resulting equivalent model can be written as
\begin{equation}
\bar{\mathbf{Y}}_k[t]=\mathbf{\Psi}_{\tau,k}\,\bar{\mathbf{W}}_k[t]+\mathbf{E}_k[t],
\label{eq:mmv_dup}
\end{equation} 
where $\mathbf{\Psi}_{\tau,k}\!\triangleq\![\mathbf{\Psi}_k\ \mathbf{\Psi}_k]$ and
$\bar{\mathbf{W}}_k[t]\triangleq\big[(\mathbf{W}^{\rm sh}_k[t])^{\top}\ \ (\mathbf{W}^{\rm pr}_k[t])^{\top}\big]^{\top}$. The corresponding TO calibration will be detailed in the next section.

Given the estimated CSI $\{\mathbf{Y}_k[t]\}_{t=1}^{T}$, we aim to recover the sparse coefficient matrix $\{\bar{\mathbf{W}}_k[t]\}_{t=1}^{T}$ after TO removal, whose nonzero support identifies the active delay-grid indices, while the associated coefficients absorb the Doppler/CFO/PO and AoA information. Hence, under \eqref{eq:mmv_dup}, the problem reduces to support recovery and coefficient estimation, from which the remaining sensing parameters can be inferred.

\section{Sensing Parameter Estimation Scheme}
In this section, we develop an efficient  estimation scheme for OFDMA multi-user uplink sensing with delay-cluster commonality under clock asynchronism. As detailed below, we estimate and compensate the TO using the known geometric LoS delay reference.
Then, leveraging the delay-clustered multi-user structure, we perform delay-clustering SBL to recover the active delay support. Finally, based on the recovered delay taps, CFO/PO are canceled using the LoS reference tap. Doppler is then estimated from adjacent-packet phase evolution, while AoA is obtained from adjacent-antenna phase differences and the path gain is recovered from the corresponding averaged coefficient correlation.
\subsection{Delay-Clustering SBL}

To recover $\{\bar{\mathbf{W}}_k[t]\}_{t=1}^{T}$ in \eqref{eq:mmv_dup}, we first need to suppress TO from $\{\mathbf{Y}_k[t]\}_{t=1}^{T}$. Since the LoS path is assumed to be dominant, the observed LoS delay can be identified from the dominant peak of the delay-domain periodogram: for packet $t$, we form
$p_k(\tau_g;t)\triangleq \frac{1}{M}\|\mathbf{d}_k^H(\tau_g)\mathbf{Y}_k[t]\|_2^2$,
with $\mathbf{d}_k(\tau_g)\triangleq [e^{-j2\pi f_n\tau_g}]_{n\in\mathcal{N}_k}$.
The strongest grid peak yields a coarse LoS delay estimate, which is further refined by the parabolic interpolation around the dominant peak to obtain $\tilde{\tau}_{k,0}^{\rm obs}[t]$ \cite{753016}. We then estimate the TO as
$\widehat{\delta\tau}_k[t]=\tilde{\tau}_{k,0}^{\rm obs}[t]-\tau_{k,0}^{\rm geom}$,
and compensate it by
$\bar{\mathbf{Y}}_k[t]=\mathrm{diag}(\boldsymbol{\phi}_k[t])\mathbf{Y}_k[t]$,
where $\boldsymbol{\phi}_k[t]\triangleq [e^{j2\pi f_n\widehat{\delta\tau}_k[t]}]_{n\in\mathcal{N}_k}$.
Above calibration mitigates the dominant TO and aligns the observations to a common delay reference for the subsequent sparse support recovery.

Then, based on \eqref{eq:W_sum_new}, to model user clustering, we consider $C$ candidate clusters and define a latent cluster indicator
$\mathbf{z}_k \triangleq [z_{k,1},\ldots,z_{k,C}]^{\top}$ for UE $k$: if UE $k$ belongs to cluster $c\in\{1,\ldots,C\}$, then $z_{k,c}=1$ and $z_{k,c'}=0$ for all $c'\neq c$, so that $\sum_{c=1}^{C} z_{k,c}=1$.
Let $\boldsymbol{\pi}=[\pi_1,\ldots,\pi_C]^{\top}$ denote the cluster-proportion vector, and assign it a Dirichlet prior
$\boldsymbol{\pi}\sim\mathrm{Dir}(\boldsymbol{\alpha}_0)$ with concentration $\boldsymbol{\alpha}_0=[\alpha_{0,1},\ldots,\alpha_{0,C}]^{\top}$.
Given $\boldsymbol{\pi}$, $\mathbf{z}_k$ follows a categorical distribution:
\begin{equation}\label{eq:prior_z}
p(\mathbf{z}_k\mid \boldsymbol{\pi})=\prod_{c=1}^{C}\pi_c^{z_{k,c}}, \qquad k=1,\ldots,K.
\end{equation}

To proceed, following the  classical sparse Bayesian model, 
we let $\mathbf{w}^{\rm sh}_{k,g}[t]$ and $\mathbf{w}^{\rm pr}_{k,g}[t]$ 
denote the $g$-th rows of $\mathbf{W}^{\rm sh}_k[t]$ and $\mathbf{W}^{\rm pr}_k[t]$ respectively, which
 admit the following Gaussian prior distributions:
\begin{align}
p\!\left(\mathbf{w}^{\rm sh}_{k,g}[t]\mid \mathbf{z}_k,\boldsymbol{\gamma}_g\right)
&= \prod_{c=1}^{C}
\mathcal{CN}\!\left(\mathbf{0},\gamma_{g,c}^{-1}\mathbf{I}_M\right)^{z_{k,c}}
= p^{\rm sh}_{k,g}[t], \label{eq:prior_wsh}\\
p\!\left(\mathbf{w}^{\rm pr}_{k,g}[t]\mid \eta_{k,g}\right)
&= \mathcal{CN}\!\left(\mathbf{0},\eta_{k,g}^{-1}\mathbf{I}_M\right)
= p^{\rm pr}_{k,g}[t]. \label{eq:prior_wpr}
\end{align}
where $\boldsymbol{\gamma}_g \triangleq [\gamma_{g,1},\ldots,\gamma_{g,C}]^{\top}$, 
$\{\gamma_{g,c}\}$ are shared row precisions governing the common support within cluster $c$,
and $\{\eta_{k,g}\}$ are user $k$-specific row precisions capturing individual components. 
Given $\{\gamma_{g,c}\}$ and $\{\eta_{k,g}\}$, all rows are assumed to be independent, yielding
\begin{equation}\label{eq:prior_W}
p\!\left(\bar{\mathbf{W}}_k[t]\mid \mathbf{z}_k,\{\boldsymbol{\gamma}_g\}_{g=1}^{G},\{\eta_{k,g}\}_{g=1}^{G}\right)
= \prod_{g=1}^{G} p^{\rm sh}_{k,g}[t]\, p^{\rm pr}_{k,g}[t].
\end{equation}

To enable conjugate variational updates, we impose Gamma hyperpriors to the precisions as
\begin{equation}\label{eq:prior_gamma_eta}
p(\gamma_{g,c})=\Gamma(\gamma_{g,c}\mid a_0,b_0),
p(\eta_{k,g})=\Gamma(\eta_{k,g}\mid a_0,b_0),
\end{equation}
where $\Gamma(x\mid a_0,b_0)$ denotes the Gamma density with shape $a_0$ and rate $b_0$.
In this work, we set $a_0=b_0=0.01$. 
Moreover, the additive noise is modeled as i.i.d. circularly symmetric complex Gaussian:
$\mathrm{vec}(\mathbf{E}_k[t])\sim\mathcal{CN}(\mathbf{0},\beta^{-1}\mathbf{I})$
with $p(\beta)=\Gamma(\beta\mid a_0,b_0)$.

Based on $\bar{\mathbf{Y}}_k[t]$, 
the conditional likelihood is given by the circularly symmetric complex Gaussian distribution:
\begin{equation}
p(\bar{\mathbf Y}_k[t]\!\mid\!\bar{\mathbf W}_k[t],\beta)
\propto
\exp\!\Big(
-\beta\big\|\bar{\mathbf Y}_k[t]-\mathbf\Psi_{\tau,k}\bar{\mathbf W}_k[t]\big\|_F^2
\Big).
\label{eq:likelihood}
\end{equation}

Furthermore, to enhance the robustness of delay-domain clustering, we leverage multiple packets and form a stacked multiple measurement vector (MMV) observation. 
Namely, the CSI estimation can be stacked as
$\tilde{\mathbf{Y}}_k \triangleq [\bar{\mathbf Y}_k[t_1],\ldots,\bar{\mathbf Y}_k[t_T]] \in \mathbb{C}^{N_k\times MT}$ and the stacked coefficient matrix can be denoted as
$\tilde{\mathbf W}_k \triangleq [\bar{\mathbf W}_k[t_1],\ldots,\bar{\mathbf W}_k[t_T]]$.

Let us define all the parameters to be estimated as
$\Omega \triangleq \Big\{\tilde{\mathbf{W}}_k\}_{k=1}^{K},\ \{\gamma_{g,c}\}_{g=1,c=1}^{G,C},\ \{\eta_{k,g}\}_{k=1,g=1}^{K,G},\ \{\mathbf z_k\}_{k=1}^{K},\ \boldsymbol{\pi},\ \beta \Big\}.$ Nevertheless, the posterior $p(\Omega\mid\{\tilde{\mathbf{Y}}_k\})$ is analytically intractable due to  $\{\mathbf z_k\}$ and $\{\gamma_{g,c}\}$.
 To address this, we adopt a mean-field variational inference (VI) approach following \cite{8552436}, which can approximate $p(\Omega\mid\{\tilde{\mathbf Y}_k\})$ by the factorized distribution $q(\Omega)$:
\begin{equation}
q(\Omega)=\prod_{k=1}^{K} q(\tilde{\mathbf W}_k)\prod_{g,c} q(\gamma_{g,c})\prod_{k,g} q(\eta_{k,g})
\prod_{k=1}^{K} q(\mathbf z_k)\,q(\boldsymbol{\pi})\,q(\beta).
\label{eq:mf_factor}
\end{equation}
According to \cite{8552436}, the variational factors in \eqref{eq:mf_factor} can be iteratively updated by coordinate ascent as
$\log q(x)\propto \mathbb{E}_{q(\Omega\setminus x)}[\log p(\{\tilde{\mathbf{Y}}_k\},\Omega)]$, 
which yields closed-form updates as summarized below.

\textit{a) Update of $q(\tilde{\mathbf{W}}_k)$}: The variational posterior of $\tilde{\mathbf{W}}_k$ is a circularly symmetric complex Gaussian matrix distribution with independent columns sharing a common row covariance, i.e.,
$q(\tilde{\mathbf{W}}_k)=\mathcal{CN}\!\left(\tilde{\mathbf{W}}_k;\mathbf U_k,\mathbf{\Sigma}^{\rm row}_k\right)$, where $\mathbf{\Sigma}^{\rm row}_k$ and $\mathbf U_k$ denote the posterior row-covariance and mean, respectively
\begin{equation}
\mathbf{\Sigma}^{\rm row}_k=\big(\mathbb{E}[\beta]\mathbf{G}_k+\mathrm{diag}(\boldsymbol{\lambda}_k)\big)^{-1},
\mathbf U_k=\mathbb{E}[\beta]\mathbf{\Sigma}^{\rm row}_k\mathbf{B}_k,
\label{eq:W_update}
\end{equation}
where $\mathbf{G}_k \triangleq \mathbf{\Psi}_{\tau,k}^{\mathsf H}\mathbf{\Psi}_{\tau,k}$,
$\mathbf{B}_k \triangleq \mathbf{\Psi}_{\tau,k}^{\mathsf H}\tilde{\mathbf{Y}}_k$ and $\boldsymbol{\lambda}_k
=\left[\bar{\boldsymbol{\gamma}}^{\rm sh}_k;\mathbb{E}[\boldsymbol{\eta}_k]\right]$, 
$\bar{\boldsymbol{\gamma}}^{\rm sh}_k
\triangleq
\big[\bar{\gamma}^{\rm sh}_{k,1},\ldots,\bar{\gamma}^{\rm sh}_{k,G}\big]^{\top}$ and $\bar{\gamma}^{\rm sh}_{k,g}
\triangleq
\sum_{c=1}^{C} r_{k,c}\,\mathbb{E}[\gamma_{g,c}]$; $\boldsymbol{\eta}_k
\triangleq
\left[
\eta_{k,1},\ldots,\eta_{k,G}
\right]^{\top}$.

Moreover, the required row-wise second moments are
$\mathbb{E}\!\left[\|\mathbf{w}^{\rm sh}_{k,g}\|_2^2\right]
=\|\mathbf U_k(g,:)\|_2^2+MT[\mathbf{\Sigma}^{\rm row}_k]_{g,g}$
and
$\mathbb{E}\!\left[\|\mathbf{w}^{\rm pr}_{k,g}\|_2^2\right]
=\|\mathbf U_k(G+g,:)\|_2^2+MT[\mathbf{\Sigma}^{\rm row}_k]_{G+g,G+g}$.

\textit{b) Update of $q(\gamma_{g,c})$ and $q(\eta_{k,g})$}:
The posteriors remain Gamma distributed:
\begin{align}
q(\gamma_{g,c})=\Gamma\!\left(a_{g,c}^{\gamma},b_{g,c}^{\gamma}\right),\;
q(\eta_{k,g})=\Gamma\!\left(a_{k,g}^{\eta},b_{k,g}^{\eta}\right).
\label{eq:gamma_eta_updates_short}
\end{align}
where $a_{g,c}^{\gamma}\!=\!a_0\!+\!MT\sum_{k=1}^{K} r_{k,c}$, $b_{g,c}^{\gamma}\!=\!b_0\!+\!\sum_{k=1}^{K} r_{k,c}\,\mathbb{E}\!\left[\|\mathbf{w}^{\rm sh}_{k,g}\|_2^2\right]$, and $a_{k,g}^{\eta}\!=\!a_0\!+\!MT$, $b_{k,g}^{\eta}\!=\!b_0\!+\!\mathbb{E}\!\left[\|\mathbf{w}^{\rm pr}_{k,g}\|_2^2\right]$.
Thus $\mathbb{E}[\gamma_{g,c}]=a_{g,c}^{\gamma}/b_{g,c}^{\gamma}$ and $\mathbb{E}[\eta_{k,g}]=a_{k,g}^{\eta}/b_{k,g}^{\eta}$.

\textit{c) Update of $q(\mathbf z_k)$ and $q(\boldsymbol{\pi})$}:
$q(\mathbf z_k)$ is categorical with responsibilities $r_{k,c}\triangleq q(z_{k,c}=1)$, updated as
\begin{equation}
r_{k,c}=\exp(\xi_{k,c})/\sum_{c'=1}^{C}\exp(\xi_{k,c'}),
\label{eq:qz_update}
\end{equation}
where $\xi_{k,c}\triangleq \mathbb{E}[\log\pi_c]
+MT\sum_{g=1}^{G}\mathbb{E}[\log\gamma_{g,c}]
-\sum_{g=1}^{G}\mathbb{E}[\gamma_{g,c}]\,\mathbb{E}\!\left[\|\mathbf{w}^{\rm sh}_{k,g}\|_2^2\right]$, $\mathbb{E}[\log \pi_c]=\psi(\alpha_c)-\psi\!\left(\sum_{c'}\alpha_{c'}\right)$, $\mathbb{E}[\log \gamma_{g,c}]=
\psi(a_{g,c}^{\gamma})-\log b_{g,c}^{\gamma}$, where $\psi(\cdot)$ denotes the digamma function.

Given a Dirichlet prior $p(\boldsymbol{\pi})=\mathrm{Dir}(\boldsymbol{\alpha}_0)$, we have $q(\boldsymbol{\pi})=\mathrm{Dir}(\boldsymbol{\alpha})$ with $\alpha_c=\alpha_{0,c}+\sum_{k=1}^{K} r_{k,c}$.

\textit{d) Update of $q(\beta)$}:
The variational posterior is $q(\beta) = \Gamma(a_\beta^{\rm post},b_\beta^{\rm post})$, where $a_\beta^{\rm post}=a_0 + M\!T\sum_{k=1}^{K} N_k$, $b_\beta^{\rm post}= b_0 + \sum_{k=1}^{K}\!\Big(
\big\|\tilde{\mathbf{Y}}_k-\mathbf{\Psi}_{\tau,k}\mathbf U_k\big\|_F^2
+ M\!T\,\operatorname{tr}\!\big(\mathbf{\Sigma}_k^{\rm row}\mathbf{G}_k\big)
\Big)$
and $\mathbb{E}[\beta]=a_\beta^{\rm post}/b_\beta^{\rm post}$.

See Algorithm~\ref{alg:1} for a summary of above proposed procedure, where $\widehat{\mathcal I}_k$
denotes the estimated active delay index set for UE~$k$, with $\widehat{L}_k=|\widehat{\mathcal I}_k|$.

\subsection{Doppler, AoA and Channel Gains Estimation}
Given the estimated active delay index set $\widehat{\mathcal I}_k$ in Section~III-A, we next reconstruct the per-tap coefficients and extract Doppler/AoA for UE~$k$.

Before the fixed-support reconstruction, we mitigate the off-grid error of the coarse delay grid via a coarse-to-fine refinement: for each detected coarse-grid tap $\hat{\tau}_{k,\ell}$, we construct a fine grid $\mathcal{T}_{k,\ell} \triangleq \left\{\hat{\tau}_{k,\ell} + i\frac{\Delta\tau}{F}\; \big| \; i=-I,\ldots,I \right\}$,
where $\Delta\tau=\frac{\tau_{\max}}{G-1}$ is the coarse-grid spacing, $F=4$ denotes the refinement factor, and $I=8$ specifies the neighborhood span.
For each candidate $\tau\in\mathcal{T}_{k,\ell}$, we compute the delay steering vector
$\boldsymbol{d}_k(\tau)=\exp(-j2\pi \mathbf{f}_{\mathcal{N}_k}\tau)\in\mathbb{C}^{N_k}$,
and refine the delay by maximizing a local delay-domain periodogram over $\tilde{\mathbf{Y}}_k$, i.e., $\bar{\tau}_{k,\ell} \triangleq \arg\max_{\tau\in\mathcal{T}_{k,\ell}}
\left\|\boldsymbol{d}_k^{\mathsf{H}}(\tau)\,\tilde{\mathbf{Y}}_k\right\|_2^2$.

Define $\bar{\boldsymbol{\tau}}_k\triangleq[\bar{\tau}_{k,1},\ldots,\bar{\tau}_{k,\widehat L_k}]^{\top}$.
Accordingly, the refined reduced delay dictionary is given by
$\mathbf B_k \triangleq [\boldsymbol d_k(\bar{\tau}_{k,1}),\ldots,\boldsymbol d_k(\bar{\tau}_{k,\widehat L_k})]\in\mathbb C^{N_k\times \widehat L_k}$
and thus the per-packet CSI estimation admits the fixed-support model 
\begin{equation}\label{eq:fixed_support_refined}
\bar{\mathbf Y}_k[t] \approx \mathbf B_k\,\widehat{\mathbf X}_k[t] + \mathbf E_k[t],
\end{equation}
where $\widehat{\mathbf X}_k[t]\in\mathbb C^{\widehat L_k\times M}$ collects the per-tap complex coefficients across the $M$ antennas.

Based on \eqref{eq:fixed_support_refined},  we next extract $\widehat{\mathbf X}_k[t]$ for each packet $t$ via a regularized least-squares (LS) projection as
\begin{equation}\label{eq:W_LS_refined}
\widehat{\mathbf X}_k[t]
=
\big(\mathbf B_k^{\mathsf H}\mathbf B_k+\lambda\mathbf I \big)^{-1}
\mathbf B_k^{\mathsf H}\bar{\mathbf Y}_k[t],
\end{equation}
where $\lambda=10^{-3}$, and the $(\ell,m)$-th entry of $\widehat{\mathbf X}_k[t]$ approximately follows the slow-time phase model $\hat x_{k,\ell,m}[t]\approx \alpha_{k,\ell}\,[\mathbf a(\theta_{k,\ell})]_m\exp\!\Big(j\beta_k[t]\Big)\exp\!\Big(j2\pi tT_A\big(\nu_{k,\ell}+\delta \nu_k[t]\big)\Big)$.

Next, in order to estimate the Doppler frequency, we construct the adjacent-packet conjugate product $\hat x^{*}_{k,\ell,m}[t]\hat x_{k,\ell,m}[t{+}1]\approx\allowbreak
|\alpha_{k,\ell}|^2|a(\theta_{k,\ell})_m|^2\,
e^{j(\beta_k[t{+}1]-\beta_k[t])}\allowbreak
e^{j2\pi T_A(\nu_{k,\ell}+(t{+}1)\delta \nu_k[t{+}1]-t\delta \nu_k[t])}$.
Exploiting the known LoS reference tap ($\nu_{k,\ell_{\rm ref}}=0$), we multiply the above product by its conjugated LoS counterpart
so that the common CFO/PO terms are canceled and we can obtain the Doppler estimate as
\begin{equation}\label{eq:20}
\begin{aligned}
\widehat{\nu}_{k,\ell}
=\frac{1}{2\pi T_A}\angle\!\Bigg(
\sum_{t=1}^{T-1}
\Big(\sum_{m=1}^{M}\hat x_{k,\ell_{\rm ref},m}[t]\hat x^{*}_{k,\ell_{\rm ref},m}[t{+}1]\Big)
\\[-1mm]
\hspace{18mm}\times
\Big(\sum_{m=1}^{M}\hat x^{*}_{k,\ell,m}[t]\hat x_{k,\ell,m}[t{+}1]\Big)
\Bigg).
\end{aligned}
\end{equation}

\begin{algorithm}[t]
\caption{Mean-field VI for cluster delay-SBL}
\label{alg:1}
\begin{algorithmic}[1]
\REQUIRE $\{\mathbf Y_k[t]\}_{t\in T}$, $\tau^{\rm geom}_{k,0}$, $\mathbf{\Psi}_{\tau,k}$.
\STATE \textbf{Prepare:} Obtain TO-compensated $\{\bar{\mathbf Y}_k[t]\}_{t\in T}$ and stack $T$ packets to form $\tilde{\mathbf{Y}}_k\in\mathbb C^{N_k\times MT}$.
\STATE \textbf{Initialize} $\{r_{k,c}\}$, $\{\mathbb E[\gamma_{g,c}]\}$, $\{\mathbb E[\eta_{k,g}]\}$, and $\mathbb E[\beta]$.
\REPEAT
\STATE Update $q(\tilde{\mathbf W}_k)$ via \textit{a)}.
\STATE Update $q(\gamma_{g,c})$ and $q(\eta_{k,g})$ via \textit{b)}.
\STATE Update $q(\boldsymbol{\pi})$ and $q(\mathbf z_k)$ via \textit{c)}.
\STATE Update $q(\beta)$ via \textit{d)}.
\UNTIL convergence
\STATE \textbf{Output:} $\{\widehat{\mathcal I}_k\}_{k=1}^K$,  where the delay support $\hat{\mathcal I}_{k}$ is obtained by selecting the dominant peaks of $\|\mathbf U_k(g,:)\|_2^2+\|\mathbf U_k(G+g,:)\|_2^2$,
yielding $\{\hat{\tau}_{k,\ell}\}_{\ell=1}^{\widehat L_k}$.
\end{algorithmic}
\end{algorithm}

Finally, since the adjacent-antenna conjugate product of the recovered $\ell$-th tap satisfies
$\hat x_{k,\ell,m}[t]\hat x_{k,\ell,m+1}^{*}[t]\approx |\alpha_{k,\ell}|^{2}e^{j\pi\sin(\theta_{k,\ell})}$,
we can derive the AoA and channel gain of the $\ell$-th path by averaging the cross-correlations between any two adjacent antenna elements averaged over $t$ and $m$ as
\begin{equation}\label{eq:aoa_sin}
\sin(\widehat{\theta}_{k,\ell})\approx
\frac{1}{\pi}\angle\!\left(
\frac{1}{T(M-1)}\sum_{t=1}^{T}\sum_{m=1}^{M-1}
\hat x_{k,\ell,m}[t]\hat x_{k,\ell,m+1}^{*}[t]
\right),
\end{equation}
\begin{equation}\label{eq:gain_hat}
|\widehat{\alpha}_{k,\ell}|^2\approx
\left|
\frac{1}{T(M-1)}\sum_{t=1}^{T}\sum_{m=1}^{M-1}
\hat x_{k,\ell,m}[t]\hat x_{k,\ell,m+1}^{*}[t]
\right|.
\end{equation}

\section{Simulation Results}
In this section, we evaluate the proposed estimation pipeline for OFDMA uplink sensing under asynchronous transceivers. 
Unless otherwise specified, the system operates at a carrier frequency of $f_c=3.5$~GHz, with a signal bandwidth of $B=140$~MHz. The total number of subcarriers is $N=2048$, 
the subcarrier spacing is $\Delta f=60$~kHz, 
and the packet interval is
$T_A\approx 0.25~\mathrm{ms}$. 
We consider a BS equipped with a $M=8$-antenna ULA with half-wavelength spacing, serving $K=8$ uplink UEs over $T=16$ packets.
The $K$ UEs are partitioned into $S=3$ delay clusters, where UEs within the same cluster share a set of common delay taps. 
Specifically, each UE contains $L_{\mathrm{sh}}=3$ commonly shared paths and $L_{\mathrm{pr}}=1$ private path (set as the LoS tap), $L_k = L_{sh} +L_{pr}$. 
The maximum delay is $\tau_{\max}=2.5~\mu$s and the grid number is $G=256$.


The Doppler frequencies are assumed to lie in $[-0.35,0.35]$~kHz, and the AoAs in $[-90^\circ,90^\circ]$. The TOs and CFOs are uniformly drawn from $[0~\text{s},\,20/B]$ and $[0~\text{Hz},\,150~\text{Hz}]$, respectively.

Before presenting the simulation results, we define the performance metrics averaged over all $K$ UEs: for $x\in\{\tau,\nu\}$, $\mathrm{NMSE}(x)\triangleq \frac{1}{K}\sum_{k=1}^{K}
\frac{\sum_{\ell\neq \ell_{\rm LoS}}\big(\widehat{x}_{k,\ell}-x_{k,\ell}\big)^2}
{\sum_{\ell\neq \ell_{\rm LoS}}x_{k,\ell}^2}$, and for AoA $\theta$ we have $\mathrm{RMSE}(\theta)\triangleq \frac{1}{K}\sum_{k=1}^{K}
\sqrt{\frac{1}{L_k}\sum_{\ell=1}^{L_k}\big(\widehat{\theta}_{k,\ell}-\theta_{k,\ell}\big)^2}$.

In Fig.~2, we evaluate the estimation performance versus SNR for two schemes: our proposed \emph{multi-user SBL} and the \emph{Individual-SBL} baseline (standard per-UE SBL \cite{tipping2001sparse} without exploiting cross-user structure).
From Fig.~2, it can be observed that the NMSE/RMSE metrics decrease monotonically with SNR. More importantly, the proposed scheme achieves uniformly lower delay/Doppler NMSE and AoA RMSE than the Individual-SBL baseline across the SNR range, with a more significant gain in the low-SNR regime.
To gain more insights, we record the multi-user clustering accuracy versus SNR in Table~\ref{tab:group_acc_snr}. As we can see, the clustering accuracy increases steadily with SNR, indicating that the proposed method can identify the underlying delay clusters more reliably as the observation quality improves.

\begin{figure}[t]
\centering
\includegraphics[bb=0 10 440 295,scale=0.35]{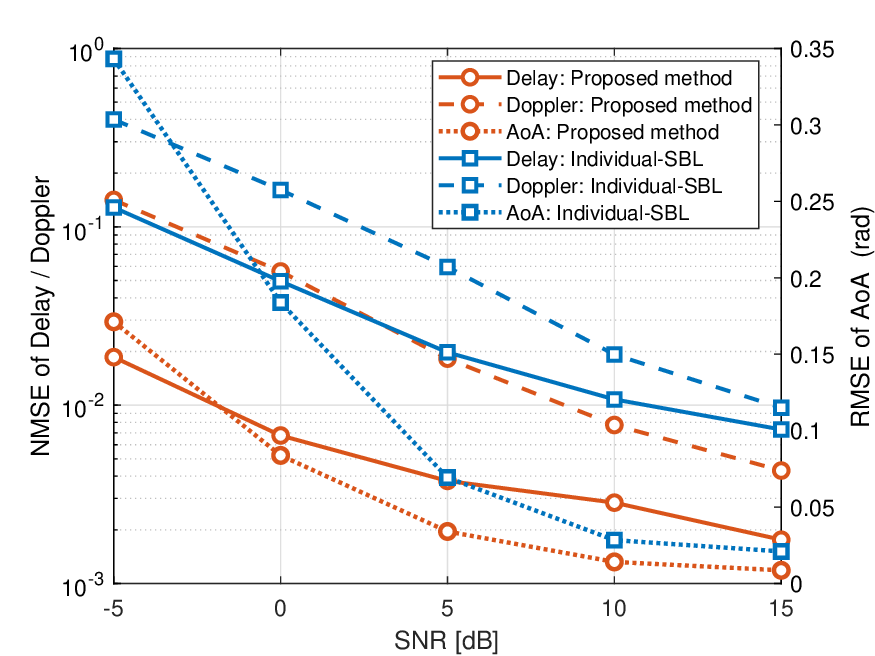}
\caption{Delay and Doppler NMSE, AoA RMSE versus SNR. $N_k=128$.}
\end{figure}

\begin{table}[t]
  \centering
  \caption{Multi-user clustering accuracy versus SNR.}
  \label{tab:group_acc_snr}
  \vspace{-1mm}
  \begin{tabular}{c c c c c c}
    \toprule
    SNR (dB) & -5 & 0 & 5 & 10 & 15 \\
    \midrule
    Clustering accuracy & 0.502 & 0.689 & 0.812 & 0.868 & 0.899 \\
    \bottomrule
  \end{tabular}
  \vspace{-2mm}
\end{table}

\begin{figure}[t]
\centering
\includegraphics[bb=0 10 440 295,scale=0.35]{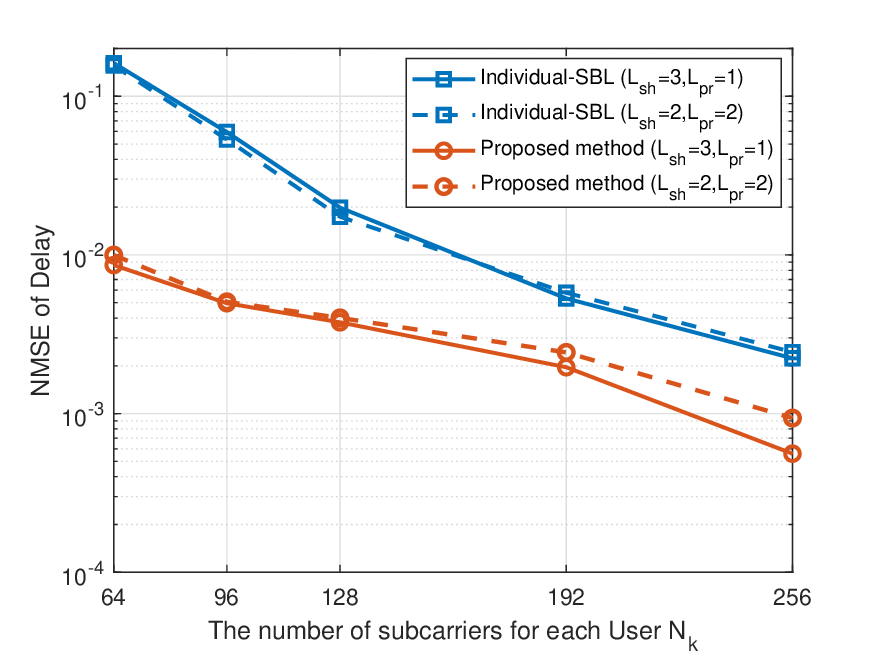}
\caption{Delay NMSE versus $N_k$. SNR = 5 dB.}
\end{figure}

\begin{figure}[t]
\centering
\includegraphics[bb=0 10 440 295,scale=0.35]{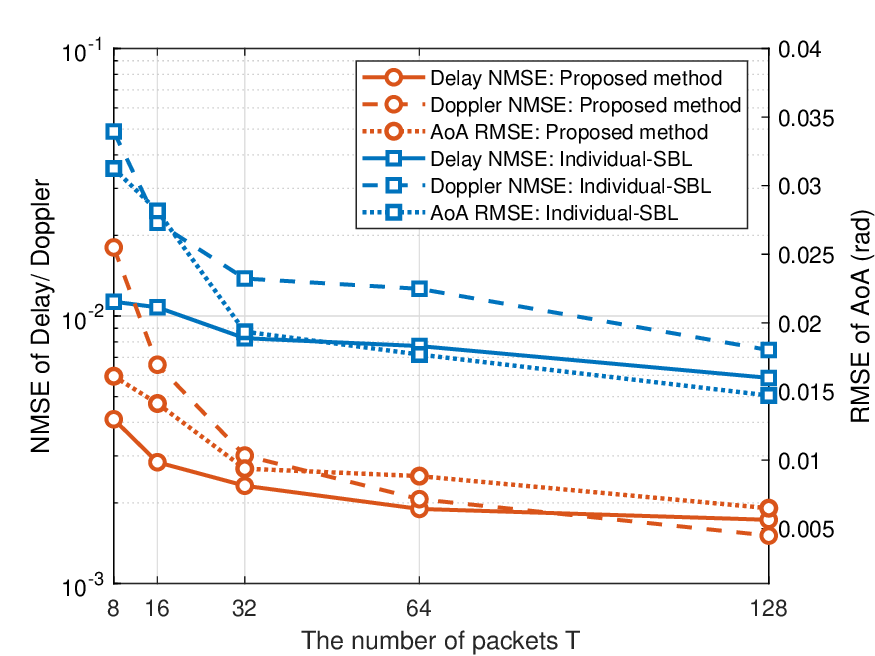}
\caption{Delay/ Doppler NMSE and AoA RMSE versus $T$. SNR = 10 dB, $N_k=128$.}
\end{figure}
Fig.~3 depicts the delay-estimation NMSE versus the number of allocated subcarriers $N_k$ under 
$(L_{\mathrm{sh}},L_{\mathrm{pr}})\in\{(3,1), (2,2)\}$.
As $N_k$ increases, the delay NMSE of all schemes decreases due to richer frequency-domain observations.
Notably, the Individual-SBL baseline is largely insensitive to the shared/private composition, since it processes each UE independently.
In contrast, the proposed multi-user scheme benefits more from stronger shared sparsity, and thus performs better in the $(3,1)$ case at large $N_k$.
Furthermore, the proposed multi-user scheme achieves a clear NMSE reduction over the Individual-SBL baseline, with a
more significant gain when $N_k$ is small, indicating exploiting cross-user shared sparsity is most beneficial in the $N_k$-limited regime.


Finally, Fig.~4 reports the delay NMSE, Doppler NMSE, and AoA RMSE versus the number of packets $T$. As $T$ increases, all three metrics decrease, as more packets provide richer temporal observations for delay support recovery, enhance slow-time diversity for Doppler estimation, and enable more effective coherent averaging for AoA estimation. Meanwhile, the proposed scheme consistently outperforms the Individual-SBL baseline across all $T$. This gain mainly stems from the improved delay support recovery enabled by exploiting common delay-cluster sparsity across users, which yields lower delay NMSE and subsequently leads to better Doppler and AoA estimation compared to Individual-SBL.
\section{Conclusions}
In this correspondence, we investigated asynchronous uplink networked sensing by exploring multiuser correlation and diversity in OFDMA-based PMNs under a novel bistatic multi-transmitter single-receiver topology. Leveraging LoS-referenced calibration and a shared--private delay-cluster SBL formulation, the proposed framework jointly exploited cross-user delay correlation and mitigated uplink asynchronism for reliable delay support recovery. Based on the recovered delay taps, Doppler and AoA were subsequently estimated from temporal and spatial phase differences. Simulation results verified consistent sensing gains over per-user processing, particularly at low-SNR and limited per-UE subcarrier budgets.


\vspace{-0.5em}
\setlength{\baselineskip}{12pt}
\bibliographystyle{IEEEtran}
\bibliography{Reference}
\end{document}